\documentclass[prl,twocolumn,amsmath,amssymb]{revtex4}
\usepackage{dcolumn}
\usepackage{bm}
\usepackage{graphicx}
\begin{document}
\title{Magnetic Glass formed by kinetic arrest of first order phase transitions}
\author{Praveen Chaddah and Alok Banerjee}
\affiliation{UGC-DAE Consortium for Scientific Research, \\University Campus, Khandwa Road,
Indore-452001, M.P, India}

\begin{abstract}
Metallic glasses are formed by splat-cooling; this ensures that atomic motions are arrested \textit{before} the latent heat of solidification can be extracted. Glass is defined as a higher disorder metastable state with arrested kinetics. Arrested kinetics is a defining property of a glass, rather than structural disorder as, e.g., in amorphous silicon \cite{greer}. `Magnetic glasses' identified recently possess structural as well as magnetic long-range order, but show relaxation and specific heat as for a structural glass\cite{greer}. The values of T$_C$ and T$_g$ in these materials are easily varied by varying the magnetic field H, allowing one to go from metallic glass to glass former like scenario.
\end{abstract}

\maketitle
\section {Introduction}
There has been a resurgence of interest in first order phase transitions, with magnetic field induced transitions providing the impetus because of possible applications envisaged for magnetocaloric materials, for materials showing large magnetoresistance, for magnetic shape memory alloys, etc. The control variables, magnetic field (H) and temperature (T) can be varied much more easily (here we do not require a medium to control H) than the control variables pressure (P) and T (here a medium is required to control P); resulting in some very interesting observations on metastable states. 

First order phase transitions are defined by a discontinuous change in entropy at T$_C$ (resulting in a latent heat) and a discontinuous change in either volume or magnetization (depending on whether Tc changes with pressure or with magnetic field). Since liquids and solids have different densities, the crystallization of a liquid entails motion at a molecular level that requires non-zero time. The concept of rapidly freezing a liquid out of equilibrium has been exploited for producing splat-cooled metallic glasses \cite{greer} whose density is that of the liquid but where `flow' also occurs over astronomical time scales. 
  
\maketitle
\section {Freezing on heating: disorder is not necessarily structural}
We associate higher entropy with more disorder, and are intuitively comfortable with the liquid being more disordered than the crystal, and the vapour being more disordered than the liquid. This disorder is in real space in the positions of the molecules, and as we are taught, it should be observed through x-ray diffraction. But if we take liquid helium-3 at a temperature below 0.1 K, subject it to a pressure of about 30 atmospheres and then start heating it, we find that the liquid freezes into a bcc solid as it is heated to about 0.2K. The solid is structurally more ordered than the liquid, and x-ray diffraction does not unravel its disorder or higher entropy. The liquid is highly ordered in momentum space (it is a quantum Fermi liquid) while the spins in solid $^3$He exhibit thermal disorder. It has been shown \cite{lee} that by adding $^4$He, this freezing on heating can be observed at even higher temperatures (upto 0.7 K) at lower pressures (about 23 atmospheres). The quantum nature of the heliums had to be invoked to understand why the solids have higher entropy than the corresponding liquids and to explain this counterintuitive first order transition of melting on cooling. We emphasize here that thermodynamics requires that the phase that exists at higher T has higher entropy or higher disorder. It does not necessarily require that it has higher structural (or real space) disorder. We emphasize this very important point that disorder can be other than structural; higher disorder does not necessarily imply higher structural (or real space) disorder

Thus if one observes a temperature-induced first order transition then thermodynamics ensures that the higher-T phase is more disordered. There are many temperature-induced first order magnetic transitions where either phase exhibits long range structural order as well as magnetic order. As discussed for solid heliums, the nature of disorder in the high-temperature phase is not structural. The persistence of the higher entropy phase below T$_C$, as a metastable supercooled state, leads to well-recognized hysteresis. We shall recognize in a later section that these can, under some conditions, smoothly change character on further cooling and persist as a metastable glass-like state. The identification of this long-range-ordered state as a ``magnetic glass" (as distinct from a supercooled state) was initially accepted about five years back \cite{chat}. 
  
\maketitle\section{Supercooling and Superheating}
First order phase transitions are defined by a discontinuous change in entropy at T$_C$ (resulting in a latent heat) and a discontinuous change in either volume or magnetization (depending on whether T$_C$ changes with pressure or with magnetic field). The existence of a latent heat implies a gradual change in the fraction of transformed phase and also that temperature cannot change from T$_C$ until the transformation is complete. As stated earlier, the coexistence of two phases at T$_C$ thus follows as an observation for identifying a first order transition. Formally stated, the two phases (say `solid' and `liquid') have different ``order parameters", but have the same free energy f at T$_C$. For a first order melting transition, this also implies that f$_{solid}$ $<$ f$_{liquid}$ at $T<T_C$; and f$_{solid}$ $>$ f$_{liquid}$ at $T > T_C$. This is formalized \cite{chaik}, following the Landau theory, by writing the free energy as 

\begin{eqnarray}
	f(T,S) = a (T-T^*)S^2 - wS^3 + uS^4
\end{eqnarray}

Here S is the order parameter, which is zero for the liquid (disordered phase); and a, w and u are positive constants independent of temperature. When plotted against S, the free energy has two minima, separated by a barrier f$_B$, for T* $<$ T $<$ [T* + 9W$^2$/16ua]. The two minima have the same value of f at $T = T_C = T^*+ w^2/2ua$. One of these minima is located at S=0; the other corresponding to the crystal is at S=S$_o$(T), the dependence on T indicating that the order in the solid rises as T falls. This is depicted in figure 1, which we use to discuss the generic evolution of a first order transition (other functional forms of f (T, S) are also used to describe first order transitions and the discussion below is valid for all those forms also).

At high T we have a liquid, with S=0, and a second minimum in f(S) starts forming at T** as T is lowered towards T$_C$. This second minimum corresponds to a possible crystalline state and at T=T$_C$ the free energies of the liquid and crystal become equal. At lower T the free energy of the crystal is lower and at much lower T the minimum in f corresponding to the liquid no longer exists. In thermodynamic equilibrium the system must have the lowest value of f and only the liquid would exist at T $>$ T$_C$ , and only the crystal would exist at T $<$ T$_C$. Both the phases would coexist at T$_C$ and, under a constant cooling rate, they would coexist until we have extracted the latent heat of the entire mass. The time $\tau$$_1$ required to extract the specific heat is dictated by the cooling rate and appears to be in our control. During the process of crystallization the density also has to change and this removal of latent heat involves motion of the molecules. This requires a time $\tau$$_2$ that is not in our control (except that $\tau$$_2$ rises as T is lowered since diffusivity falls). Usually $\tau$$_1$ $>>$ $\tau$$_2$ and we assume this here. What happens if $\tau$$_1$ $<$ $\tau$$_2$ ? This is a very interesting possibility that was consciously exploited about half-a-century back to obtain metallic glasses\cite{greer}; we shall discuss it in the next section. Metallic glasses are obtained if specific heat can be extracted without extracting latent heat i.e. by making $\tau$$_2$ $>>$ $\tau$$_1$. Thus ``extract specific heat without extracting latent heat" serves as a procedural definition of glass formation.
	
If there is no source for the energy required to cross f$_B$, then the liquid will cool below T$_C$ without crossing the barrier f$_B$. We will have a "supercooled liquid" at T $<$ T$_C$ which sits in a local minimum of f(S), and which decays to the global minimum at S$_o$(T) at a rate dictated by exp [-f$_B$(T)/kT]. We note that f$_B$(T) falls with decreasing T so that the decay of the supercooled liquid to the stable crystal is faster as T decreases. This is a crucial feature of a supercooled state. The barrier f$_B$ reduces to zero at T = T* and this is the limit to which a high-T disordered phase can be supercooled. There is no restriction on the kinetics of the supercooled system; it should not receive a fluctuation that would enable formation of a critical nucleus of the ordered phase. Careful experiments have established that T* for water is about $-40^{o}C$.

	Ice could similarly be superheated to T** but this has not been experimentally possible because the critical size for nucleation of water (any liquid) on the surface of ice (or the crystal) approaches zero. A solid has to be coated with another solid of higher melting point to prevent surface nucleation, and superheating is then observed. Water can be superheated and can remain a liquid to about $280^{o}C$. The microwave oven provides an excellent source of heating pure water without inducing fluctuations; we provide that fluctuation f$_B$ when we handle the cup of water, causing it to explode into steam. 

	We have discussed here supercooled states obtained by varying only T. More interesting results are obtained by varying the other control variable that could be pressure. The freezing point of water drops from $0^{o}C$ to about $-20^{o}C$ under increasing pressure. There are some reports of cooling water to below $0^{o}C$ under pressure, and then lowering pressure to obtain supercooled water. These experiments are not straightforward, but it is routine to cool under magnetic field, or to cool under zero field and then apply H. Many magnetic first order transitions lower magnetization with reducing temperature and these have their T$_C$ drop, with experimentally accessible H, to a small fraction of their zero-field value. The extensively studied examples are vortex matter phase transitions, charge ordering in half-doped manganites, and various FM to AFM transitions. It was argued that following different paths in the two control variable space yields different limits to which supercooling is observed \cite{pc}. 

\maketitle\section{Supercooled vs. glasslike states}
In the above discussion we have assumed time scales that allow thermodynamic free energy to determine the state of the system. In that case no metastable state exists for T $<$ T* because the barrier between the states S=0 and S=S$_o$ is now zero and the system would transform in a time $\tau$$_2$ dictated by its kinetics. We have emphasized earlier that $\tau$$_2$ increases sharply as T is lowered and below a temperature T$_g$ it would exceed the lifetime (or the patience!) of the experimentalist. If we could put the sample in the S=0 state at T $<$ T$_g$, then it would not evolve from S=0 to S=S$_o$ even if f$_B$ has become zero. When S=0 corresponds to a liquid then the first order structural transition to a crystal is arrested, and we have a structurally-disordered glass. Now that we are considering general first order transitions, we refer to a metastable disordered phase at T$_1$ $<$ T$_C$, and with $\tau$$_2$(T$_1$) larger than our experimental time scale, as a glass-like arrested state (GLAS). How do we form this glass-like state?

We have not discussed the implications of the sign of [T* -T$_g$] so far. To form a glass-like state when T$_g$ $<$ T* (corresponding to metallic glass scenario) it is essential that cooling is so rapid that $\tau$$_1$ $<<$ $\tau$$_2$ (T$_C$). Since we need the sample to cool below T$_g$ before the molecules in a liquid can adjust their separation, $\tau$$_1$ depends on the total heat capacity from T$_g$ to T$_C$. To form glasses from liquid metals this requires $\tau$$_1$ be in the range of milliseconds, and the technique of splat-cooling with cooling rates upto 10$^6$ K/sec was evolved\cite{greer}.  Since the minimum cooling rate required depends on $\tau$$_2$(T$_C$), it falls if T$_C$ is lowered. Recently, Bhat et al \cite{bhat} succeeded in vitrifying liquid Ge, whose T$_C$ falls with rising pressure, by rapid cooling under high pressure. Since $\tau$$_2$ had risen with decrease in T$_C$, the glass could be formed at achievable cooling rates.

The decrease of T$_C$ with rising magnetic field is much more rapid in many magnetic transitions - it can often be reduced to nearly 0K with accessible values of H - and the idea of observing glass-like arrest of kinetics even under normal cooling rates arose naturally many years back \cite{manekar}. If there was a temperature T$_g$ in these magnetic transitions where $\tau$$_2$(T$_g$) exceeded the experimental time scale then, since T$_C$ was dropping nearly to 0K with increasing H, we can experimentally explore the scenario T$_g$ $>$ T*. This is the glass former scenario. The T$_C$ for a magnetic transition falls with increasing H when the low-T phase is antiferromagnetic (AF) and we expect the possibility of a glass-like arrested state on slow cooling at some H. This was first observed in doped CeFe$_2$, but has since been confirmed in other intermetallic materials and also in many half-doped manganites that are extensively studied for their colossal magnetoresistance. The half doped manganites have an important advantage over other materials because the conductivity changes drastically along with magnetic order across the transition. While a decrease in global magnetization of the sample can be interpreted as either reduction of moment in ferromagnetic metallic (FM-M) phase, or as part transformation of FM-M to AF-insulator (AF-I), a simultaneous measurement of conductivity provided a clear choice between the two alternatives because of the orders of magnitude changes in resistivity associated with the metal to insulator transition in the latter case.

If T$_g$ $>$ T* then the glass-like arrested state can be obtained without much ado; the sample is in a supercooled liquid state with S=0 as T is lowered below T$_C$ and would continue in the metastable state until T*. At T = T$_g$ $>$ T* the metastable state is in a local minimum of free-energy surrounded by a non-zero f$_B$; its kinetics is however arrested and it cannot adjust its order parameter, or approach the value of S where the barrier exists. The value (or very existence) of f$_B$ thus becomes irrelevant below T$_g$, and the sample will remain in the state with S=0. The experimental signature is conceptually simple. Between T$_C$ and T$_g$ ($>$T*) the sample remains in the metastable state because of f$_B$; it is able to explore the neighbourhood of S=0. The decay rate for this metastable state rises with reducing T because f$_B$ falls. Below T$_g$ the sample remains in the metastable state because of large $\tau$$_2$(T), because it requires large times to explore the neighbourhood of S=0. The decay rate for the metastable state now falls with reducing T because $\tau$$_2$(T) rises. This non-monotonic behaviour of the decay rate signifies the existence of a glass-like arrested state obtained under slow cooling \cite{pc2}.

We now discuss a defining measurement (and protocol) developed to identify the existence of a `magnetic glass'. The underlying physical phenomenon corresponds to the devitrification of glass, which has been used as evidence for metallic glasses \cite{greer}. Metallic glasses form under extremely high cooling rates and at temperatures where restructuring require extremely long times. The absence of a local minimum in f(S) or the vanishing of the barrier f$_B$ have been made irrelevant by the astronomical times required for molecular diffusion and restructuring. Slow warming allows rapid restructuring at higher temperatures, and the metallic glass devitrifies to the crystalline state. The rates of varying T during glass-formation and during devitrification are very different. Magnetic glasses, on the other hand, are obtained by slow cooling in a field H$_o$ because T$_g$ $>$ T* at H$_o$ and the slow-cooled state is still metastable (with finite f$_B$) at the temperature T$_g$ where relaxation times have already become astronomical. At some other field H$_1$ we can have T$_g$ $<$ T*. The glassy state is retained by varying H isothermally, at T $<$ T$_g$, from H$_o$ to H$_1$. Since T$_g$ $<$ T* at H$_1$, and f$_B$ is zero at T$_g$, devitrification of the glassy state is seen during slow warming  in the field H$_1$. So, both glass-formation and devitrification are seen with the same rate of varying T; but the cooling and heating are done in different (unequal) fields \cite{ab, ab2}. This protocol, of cooling and heating in unequal fields (CHUF), was invoked with the idea that $H_o >H_1$ is required if T$_C$ (and also T*) falls with rising H, because T* has to be smaller during cooling in H$_o$ and larger during warming in H$_1$. By similar arguments, $H_o < H_1$ is required if T$_C$ rises with rising H. It can be trivially argued, from naive considerations, that the former corresponds to an FM to AFM transition with lowering T, and the latter corresponds to an AFM to FM transition with lowering T.

Since devitrification to an equilibrium crystalline state has to be followed by melting as T is raised further, we observed a reentrant change in density as a metallic glass is warmed. Similarly, a re-entrant change in magnetization is seen on warming a magnetic glass under CHUF protocol, but only for the correct sign of $[H_o-H_1]$. Crystallization (or devitrification) is observed on warming, as in a metallic glass \cite{ab, ab2, kumar, rawat, rawat2, lak}. We show in figure-2 data for magnetization for a manganite sample establishing that on warming a re-entrant transition was observed for only one sign of $[H_o-H_1]$. Varying H from 3 Tesla to 4 Tesla at 5K took the sample from glass former scenario to metallic glass scenario.

The ease of availability of liquid helium, and the ease with which magnetic field can be generated and controlled, have allowed us to ask questions that can be raised for the ubiquitous structural glasses, but may be difficult to experimentally address. The understanding of structural glasses remains a challenge in spite of being the subject matter of active research for more than 30 years.

Jamming and Structural Glass formation have similarities because both are manifestations of slowing down of translational kinetics, and are frequently compared and contrasted \cite{nature}. Magnetic glass formation does involve kinetic arrest of the underlying first order transition, but is not easily visualized as similar to jamming. We have introduced in this paper the idea that glasses are formed when the heat removal process preferentially removes specific heat, without removing latent heat. (This idea distinguishes jamming, where there need not be any underlying first order transition or latent heat.) This idea is of course valid, as any new idea must be, for structural glasses where the major contribution to latent heat comes from the movement of atoms and molecules as the density also changes from that of the liquid to that of the solid. Can our idea, which is apparently looking at why glasses form very different from the `confusion principle' \cite{greer} for glass formation, be pushed further and tested in magnetic glasses? At low temperatures the specific heat is dominated by conduction electrons in metallic systems, and so is the heat conduction process. The rearrangement of magnetic order that underlies the latent heat of the magnetic first order transition can, however, be originating from itinerant electrons or from orbital electrons. The coupling of the latent heat associated with the magnetic transition, to the conduction electrons that also conduct heat, would be different in these two cases. Will the ease of glass formation be different, and would this difference be experimentally verifiable? Will this give a different way to understand the physics underlying the formation of structural glasses? 

We hope this article will motivate young researchers to work on metastabilities across first order magnetic transitions and help resolve some outstanding issues in `glass physics'.

\begin{figure*}
	\centering
		\includegraphics{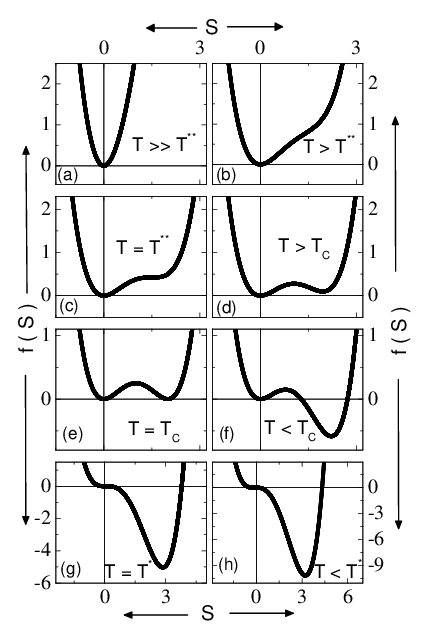}
	\caption{This schematic shows the evolution of the free-energy curves [f vs order parameter S] as temperature is varied from $T>>T_C$ to $T<<T_C$. At T corresponding to the schematics (d), (e), and (f), both the disordered and the ordered states are in a minimum of f(S), and one of these can exist as a metastable state.}
	\label{fig:Fig1}
\end{figure*}

\begin{figure*}
	\centering
		\includegraphics{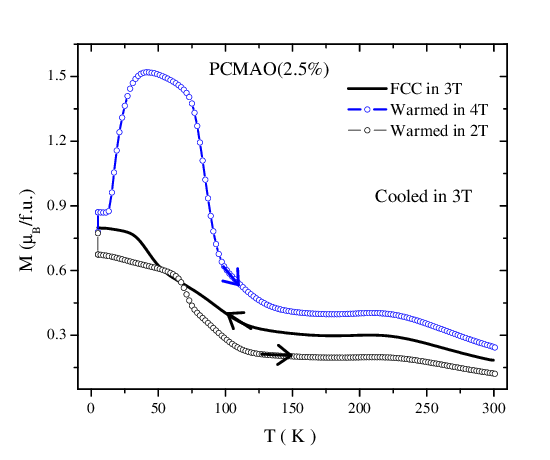}
	\caption{The starting initial state is fixed by cooling in H = 3 Tesla. On warming in the higher H = 4 Tesla one observes two sharp changes implying a reentrant transition, whereas on warming in the lower H = 2 Tesla one observes only one transition. The data is taken from Ref. [10]}
	\label{fig:Fig2}
\end{figure*}
 
\end{document}